%% file: terra-arxiv.tex
\documentclass[11pt]{article}
\usepackage[utf8]{inputenc}
\usepackage{amsmath, amssymb, amsthm, bbm, mathrsfs, mathtools}
\usepackage{hyperref}
\usepackage{geometry}
\usepackage{natbib}
\usepackage[linesnumbered,ruled,vlined]{algorithm2e}
\usepackage{verbatim}
\usepackage{multirow}
\usepackage{booktabs, array, threeparttable}
\usepackage{rotating}
\usepackage{graphicx}
\usepackage{subfigure}
\usepackage{float}
\usepackage{caption}
\usepackage{amsfonts}
\usepackage{xr}
\usepackage{enumitem}
\usepackage{xcolor}
\usepackage{comment}
\usepackage{bm}
\usepackage{tikz}
\usetikzlibrary{shapes.geometric, arrows}
\tikzstyle{startstop} = [rectangle, rounded corners, minimum width=3cm, minimum height=2cm,text centered, text width = 3cm, draw=black, fill=red!30]
\tikzstyle{io} = [trapezium, trapezium left angle=70, trapezium right angle=110, minimum width=3cm, minimum height=1cm, text centered, draw=black, fill=blue!30]
\tikzstyle{process} = [rectangle, minimum width=3.7cm, minimum height=2cm, text centered, draw=black, text width=3.5cm, fill=orange!30]
\tikzstyle{decision} = [diamond, minimum width=3cm, minimum height=1cm, text centered, draw=black, text width = 3cm, fill=green!30]
\tikzstyle{arrow} = [thick,->,>=stealth]

\mathtoolsset{showonlyrefs=true}

\usepackage{geometry}
\input{CustomizedCommands.tex}

\newcommand{\tind}[1]{\tilde{\boldsymbol{1}}_{(#1)}}
\newcommand{\indsub}[1]{\boldsymbol{1}_{(#1)}}
\newcommand{\T}{\textsc{t}}

\usepackage{xstring}
\newcommand{\showproof}[2]{%
    \IfEqCase{#1}{%
        {0}{}%
        {1}{#2}%
    }[\PackageError{showproof}{Undefined option to showproof: #1}{}]%
}%

\newtheorem{theorem}{Theorem} 
 
\newtheorem{remark}{Remark} 
 
\newtheorem{assumption}{Assumption}

\newtheorem{proposition}{Proposition}

\SetKwInput{KwInput}{Input}                % Set the Input
\SetKwInput{KwOutput}{Output}              % set the Output
\renewcommand{\theequation} {\arabic{section}.\arabic{equation}}

\allowdisplaybreaks

\title{TERRA: A Transformer-Enabled Recursive R-learner for Longitudinal Heterogeneous Treatment Effect Estimation}
\author{
Lei Shi\thanks{Lei Shi is from Adobe Research, San Jose, CA, with email: {leis@adobe.com}. Sizhu Lu is from Department of Statistics, UC Berkeley, Qiuran Lyu is from Division of Biostatistics, UC Berkeley, Peng Ding is from Department of Statistics, UC Berkeley, Nikos Vlassis is from Adobe Research.}, Sizhu Lu, Qiuran Lyu, Peng Ding, Nikos Vlassis
} 

\date{}

\begin{document}

\maketitle

\begin{abstract}
Accurately estimating heterogeneous treatment effects (HTE) in longitudinal settings is essential for personalized decision-making across healthcare, public policy, education, and digital marketing. However, time-varying interventions introduce many unique challenges, such as carryover effects, time-varying heterogeneity, and post-treatment bias, which are not addressed by standard HTE methods. To address these challenges, we introduce \textbf{TERRA} (\underline{T}ransformer-\underline{E}nabled \underline{R}ecursive \underline{R}-le\underline{A}rner), which facilitates longitudinal HTE estimation with flexible temporal modeling and learning. TERRA has two components. First, we use a Transformer architecture to encode full treatment–feature histories, enabling the representation of long-range temporal dependencies and carryover effects, hence capturing individual- and time-specific treatment effect variation more comprehensively. Second, we develop a recursive residual-learning formulation that generalizes the classical structural nested mean models (SNMMs) beyond parametric specifications, addressing post-treatment bias while reducing reliance on functional assumptions. In simulations and data applications, TERRA consistently outperforms strong baselines in HTE estimation in both accuracy and stability, highlighting the value of combining principled causal structure with high-capacity sequence models for longitudinal HTE.
    
\end{abstract}

\newpage 
\section{Introduction}

Heterogeneous treatment effect estimation (HTE) is a fundamental problem in causal inference with broad applications across healthcare \citep{kent2018personalized}, social sciences \citep{hu2023heterogeneous}, digital experimentation \citep{xie2018false}, and business decision-making, among others. The central goal is to understand how treatment impacts vary across different individuals or subgroups, moving beyond the traditional average treatment effect to identify personalized effects that can guide better interventions, policies, and decisions. HTE estimation has been well studied in the literature under different metalearner frameworks and machine learning methods \citep{wager2024causal}. 

In many real-world problems, the data structure is collected in a longitudinal manner, with repeated interventions or exposures over time. Examples include clinical trials with sequential treatment regimes, public policy programs rolled out over multiple phases, online platform experiments with evolving user interactions, and marketing campaigns with repeated customer touchpoints. In these contexts, it becomes important to consider the temporal effect in HTE estimation, i.e., the effect of a treatment at time $t$ given an individual's treatment dynamics and feature history. 

Longitudinal HTE estimation presents several challenges. (i) carryover effects: the effect of a treatment at a later timepoint may be influenced by the effect of the treatment at an earlier timepoint. (ii) time-varying effect heterogeneity: treatment effects not only vary across individuals or groups, but also vary over time. (iii) post-treatment bias: the features themselves may also be affected by past treatments, which is a common challenge in longitudinal causal inference and might lead to biased estimation. While these challenges are central to longitudinal causal inference, most existing HTE methods are not designed to address them directly.

To make these issues more concrete, we consider a few examples from several areas. In public health, the dosing and combination therapy often run over months. The effect of a dose at time $t$ depends on earlier doses and the patient's evolving condition. At the same time, important features such as biomarkers and adherence status are also affected by the trajectory of past treatment, introducing post-treatment bias if we naively control for them.
% In public policy, we are interested in the causal effects of large public programs, many of which, such as cash transfers, job training, and outreach visits are delivered in stages. Effects can accumulate or decay with the exposure history, and some intermediate outcomes (e.g., employment status) feed back into subsequent assignment, which induces confounding if we ignore them. 
% In education, treatment like adaptive tutoring adjusts content and difficulty level over sessions. The later learning is likely to depend on earlier lessons and the student's progress, yielding individual and time-specific effects. 
% In digital platforms, recommendations, notifications, and advertisements arrive as sequences. Engagement at time $t$ also depends on the past exposures and covariate shifts, leading to carryover and time-varying heterogeneity. Across these settings, attributing effects to a single intervention or ignoring temporal HTE leads to biased or unstable conclusions.
As one concrete business use case, we consider the problem of marketing attribution in tech industries. Marketing attribution targets the contribution of multiple, time‑ordered touchpoints to conversion \citep{buhalis2021bridging}. Firms interact with customers across channels over time. Common single‑touch and multi‑touch heuristics (e.g., first‑touch, last‑touch, linear, time‑decay) are ad hoc and do not address carryover, time‑varying heterogeneity, or post‑treatment bias \citep{Statsig2025, berman2018beyond, shao2011data}. We will visit this motivating scenario with a semi-synthetic study in the main paper. 

% \begin{figure}[ht!]
%     \centering
%     \includegraphics[width=0.49\textwidth]{marketing-attribution-pic.png}
%     \caption{An example of a customer journey in marketing attribution from \cite{attribution-blog-example}.}
%     \label{fig:marketing-attribution}
% \end{figure}

% edit until here...

In this work, we introduce \textbf{TERRA}, a transformer‑enabled recursive R‑learner for longitudinal HTE. TERRA employs flexible sequence modeling for heterogeneous and time-varying causal effects from full treatment–feature histories. Methodologically, we extend the structural nested mean models to a nonparametric, recursive residual‑learning formulation that addresses the aforementioned challenges and reduces reliance on functional‑form assumptions. For the architecture, we use a Transformer to capture long‑range temporal dependencies and carryover effects. Across Monte Carlo simulations and a semi-synthetic real‑world use case in marketing attribution, TERRA consistently outperforms strong baselines in accuracy and stability, and the framework is broadly applicable to real-world problems in public health, public policy, education, digital experimentation, and other business settings.

\section{Related works}

\paragraph{Meta-learners for HTE.}  HTE estimation has received extensive studies, based on various meta-learners \citep{wager2024causal}. For example, there are outcome-only approaches, such as single-learner (S-learner, \citealp{imai2013estimating}), two-learner (T-learner, \citealp{athey2016recursive}), which heavily rely on correct outcome modeling. A more popular set of learners incorporates propensity scores, such as the cross-learner (X-learner, \citealp{kunzel2019metalearners}), the doubly-robust learner (DR-learner, \citealp{kennedy2023towards}), and the residual learner (R-Learner, \citealp{nie2021quasi}). Despite these exciting progresses, the connection of these methods to a longitudinal setting is not fully explored. 

\paragraph{Longitudinal causal inference.}
% SNMMs \citep{robins1986new, robins1994correcting, robins2000sensitivity, robins2004optimal, vansteelandt2014structural, vansteelandt2016revisiting}, G-estimation \citep{robins1994correcting}, double/debiased machine learning for longitudinal data \citep{lewis2020double, battocchi2021estimating}, transformer for sequential data \citep{melnychuk2022causal, shirakawa2024longitudinal}, reinforcement learning \citep{tran2023inferring}. 
Structural nested mean models (SNMMs) provide a principled way to decompose longitudinal effects into time-indexed blips, separating incremental treatment impacts from history \citep{robins1986new,robins1994correcting}. Building on this, recent double/debiased machine-learning approaches extend orthogonalization and cross-fitting to longitudinal settings \citep{lewis2020double,battocchi2021estimating}. Our proposal is rooted in the idea of SNMMs, but provides a crucial improvement over the parametric restrictions in treatment effect modeling. A complementary line frames the problem via reinforcement learning, by constructing policies or value functions from Markov-decision processes \citep{tran2023inferring}.

\paragraph{Longitudinal counterfactual prediction based on deep sequence models.} \citet{melnychuk2022causal} proposed a Transformer-based sequence model to encode longitudinal histories and predict counterfactuals. \citet{bica2020estimating} proposed a Counterfactual Recurrent Network (CRN): a recurrent architecture that learns balanced representations over time to forecast counterfactual outcomes under different treatment sequences.
\citep{li2020g} did a neural implementation of the g-formula called G-Net that models conditional outcome dynamics. \citet{lim2018forecasting} proposed Recurrent Marginal Structural Networks (RMSN) that combine recurrent nets with inverse-propensity weighting for predicting counterfactual outcomes. All of these architectures are directly targeting the counterfactual outcome trajectory instead of the treatment effects, which differ inherently from our proposed training pipeline. 

%Deep learning based attribution models are also proposed \citep{arava2018deep, kakalejvcik2018multichannel, kumar2020camta, ren2018learning}. 

\section{Methodology}\label{sec:methodology}
\subsection{Notation and setup}
We consider a longitudinal experiment with $N$ units, each observed over $T$ time periods. At each time $t=1,\ldots,T$, the unit receives a treatment $Z_t$ and we observe features $X_t$. The final outcome of interest, denoted by $Y$, is measured at the end of the study (time $T$). The temporal structure of the observed data can be represented as
{\small 
\begin{align}\label{eqn:data-structure}
    X_0 \rightarrow Z_1 \rightarrow X_1 \rightarrow Z_2 \rightarrow \cdots \rightarrow X_{T-1} \rightarrow Z_T \rightarrow Y.
\end{align}
}
Each treatment $Z_t$ takes values in a finite set of variants $\cZ$, with one distinguished element $z^0\in \cZ$ denoting the control variant.

We adopt the potential outcome framework \citep{neyman1923application, rubin1974estimating}. For a given treatment sequence $\oz_T=(z_1,\ldots,z_T)$, we define the corresponding potential outcome of unit $i$ as $Y_i(\oz_T)$. We assume units are independently and identically distributed, so that
\begin{align*}
    Y_i(\oz_T) \sim_{i.i.d.} Y(\oz_T).
\end{align*}
We first impose the assumption that ensures the observed outcome is the same as the potential outcome corresponding to the realized treatment path.
\begin{assumption}[Outcome Consistency]
\label{cond:outcome-consistency}
    Given the realized treatment sequence $\oZ_T$, the observed outcome $Y$ equals the corresponding potential outcome $Y(\oz_T)$, i.e., $Y = Y(\oZ_T)$.
\end{assumption}

Next, we impose the standard sequential randomization assumption, which states that the treatment assignment at each period is conditionally independent of the potential outcomes, conditional on past history of treatment assignments and features.
\begin{assumption}[Sequential Randomization]
\label{cond:sequential-randomization}
    For any $t = 1, \dots, T-1$ and any treatment sequence $\oz_T$,
    \begin{align}
    \label{eqn:sequential-randomization}
        Z_{t+1} \indep Y(\oz_T) \mid \oZ_t = \oz_t, \oX_t.
    \end{align}
\end{assumption}

\subsection{Structural nested mean models (SNMMs)}
To address the challenges of longitudinal causal inference, we build on the framework of SNMMs, originally introduced by \citet{robins1986new} and further developed in \citet{robins1994correcting, robins2000sensitivity,robins2004optimal,vansteelandt2014structural,vansteelandt2016revisiting}. SNMMs are particularly well-suited to longitudinal causal inference problems because they account for carryover effects of time-varying treatments, flexibly capture heterogeneity in time-varying causal effects, and avoid the post-treatment bias described previously. 

The key idea of SNMMs is that instead of modeling the full potential outcome trajectory $Y(\oz_T)$, they model the incremental causal effect of treatment at each time point, given past treatment and feature history. This incremental causal effect is defined as the blip function:
\begin{align}
    \label{eqn:blip-function}
    & \gamma_t(\oz_t, \oX_{t-1}) 
    =\bbE \left\{ Y(\oz_t, \uz_{t+1}^0) - Y(\oz_{t-1}, \uz_t^0) \mid \oZ_t = \oz_t, \oX_{t-1} \right\}.
\end{align}
Intuitively, the blip function $\gamma_t(\oZ_t, \oX_{t-1})$ is the causal effect of receiving treatment $z_t$ at time $t$, relative to the control $z^0$, given history of treatment assignments $\oZ_t$ and the feature vector $\oX_{t-1}$. By construction, the blip function is zero when $z_t = z^0$. A common parametric representation of the blip function is
\begin{align}
    \label{eqn:blip-function-form}
    \gamma_t(\oz_t, \oX_{t-1};\theta) = \sum_{z\in \cZ^{+}} \indsub{z_t = z} g_t^z(\oz_{t-1}, \oX_{t-1};\theta),
\end{align}
where $\theta$ is the unknown parameter, $\cZ^{+}=\cZ\setminus\{z^0\}$ denotes the set of treatment levels not equal to control, and $g_t^z$ are component functions corresponding to different treatment trajectories. 
% For clarity, without loss of generality, we focus on a binary treatment setting ($\cZ = \{z^0, z^1\}$) in the following discussion.

\subsection{A residual learning framework}
\begin{figure*}[t]
    \centering
    \includegraphics[width=\textwidth]{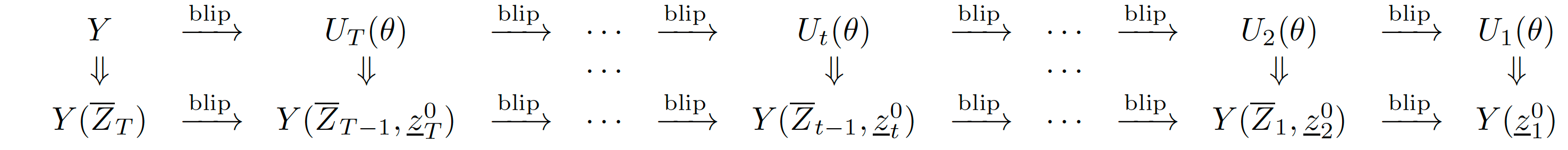}
    \caption{A diagram for the blipped potential outcome trajectory}
    \label{fig:blipped-outcome-identity-diagram}
\end{figure*}

The key insight of SNMMs is that blip functions allow us to iteratively transfer the observed outcome into a ``blipped'' outcome that mimics the counterfactual outcome under control. 
% Intuitively, as the blip functions are temporary measures of the treatment effect at each time point, we can perform a recursive blipping procedure to recover or mimic the control potential outcomes with respect to the conditional mean. 
\paragraph{Blipped outcomes.}
More specifically, for $t = T, \dots, 1$, define the recursively blipped outcomes:
\begin{align}
    U_{T+1}(\theta) = Y, \quad 
    U_{t}(\theta) = U_{t+1}(\theta) - \sum_{z\in \cZ^{+}} \indsub{Z_t = z} g_t^z(\oZ_{t-1}, \oX_{t-1};\theta).\label{eqn:blipped-outcome-recursive}
\end{align}
Equivalently,
\begin{align}
    % \label{eqn:blipped-outcome}
    U_t(\theta) = Y - \sum_{s=t}^{T} \sum_{z\in \cZ^{+}} \indsub{Z_s = z} g_s^z(\oZ_{s-1}, \oX_{s-1};\theta). 
\end{align}
If the blip functions are correctly specified, $U_t(\theta)$ behaves like the control potential outcome $Y(\oZ_{t-1}, \uz_t^0)$, conditional on $(\oZ_{t-1},\oX_{t-1})$:
\begin{proposition}[Recursive blipping property,  \cite{robins1994correcting}]
\label{prop:recursive-blipping}
For each $t=1,\ldots,T$, we have
\begin{align}
    \E{U_t(\theta) \mid \oZ_t, \oX_{t-1}}  = \E{Y(\oZ_{t-1},\uz_t^0)\mid \oZ_{t-1}, \oX_{t-1}}  = \E{U_t(\theta) \mid \oZ_{t-1}, \oX_{t-1}}. \label{eqn:blipped-outcome-identity} 
\end{align}
\end{proposition}
Intuitively, Proposition~\ref{prop:recursive-blipping} shows that the blipped outcome $U_t(\theta)$ recovers the conditional mean of the control potential outcome $Y(\oZ_{t-1}, \uz_t^0)$ given the history of treatment assignments $\oZ_{t-1}$ and the feature vector $\oX_{t-1}$. Figure~\ref{fig:blipped-outcome-identity-diagram} provides a diagram to illustrate the blipped potential outcome trajectory.

\paragraph{Estimating equation.} 
% Expanding $U_t(\theta)$ in~\eqref{eqn:blipped-outcome-identity} by plugging in the definition of $U_t(\theta)$ in~\eqref{eqn:blipped-outcome-recursive}, we have
% {\small 
% \begin{align}
%     \label{eqn:blipped-outcome-identity-expression}
%     & \bbE\Big\{U_{t+1}(\theta) - \sum_{z\in \cZ^{+}} \indsub{Z_t = z} g_t^z(\oZ_{t-1}, \oX_{t-1};\theta) \mid \oZ_{t-1}, \oX_{t-1}\Big\} \\
%     & = \mu_{t}(\oZ_{t-1}, \oX_{t-1}) - \sum_{z\in \cZ^{+}} e_t^z(\oZ_{t-1}, \oX_{t-1}) g_t^z(\oZ_{t-1}, \oX_{t-1};\theta),
% \end{align}} 
% where
% \begin{align}
%     \mu_t(\oZ_{t-1}, \oX_{t-1})
%     &= \bbE\left\{U_{t+1}(\theta) \mid \oZ_{t-1}, \oX_{t-1}\right\}, \\
%     e_t^z(\oZ_{t-1}, \oX_{t-1})
%     &= \bbP(Z_t = z \mid \oZ_{t-1}, \oX_{t-1})
% \end{align}
% denote the conditional mean of the blipped outcome $U_{t+1}(\theta)$ given the history of treatment assignments $\oZ_{t-1}$ and the feature vector $\oX_{t-1}$ and the propensity score, respectively. 

We next construct a general class of estimating equations, provided in the following Theorem.
\begin{theorem}[Estimating equations]
Under Assumptions~\ref{cond:outcome-consistency}~and~\ref{cond:sequential-randomization}, for any $t=1,\ldots,T$ and a class of measurable functions $h_t^z(\oZ_{t-1}, \oX_{t-1})$,
\begin{align}
\bbE\Big[\sum_{z\in \cZ^{+}}\tind{Z_t = z}h_t^z(\oZ_{t-1}, \oX_{t-1}) \times 
\Big\{\tU_{t+1}(\theta) - \sum_{z\in \cZ^{+}} \tind{Z_t = z} g_t^z(\oZ_{t-1}, \oX_{t-1};\theta)\Big\}\Big]=0, 
\label{eqn:blipped-outcome-estimating-equation}
\end{align}
where $\tU_{t+1}(\theta)$ and $\tind{Z_t = z}$ are the residualized version of $U_{t+1}(\theta)$ and $\indsub{Z_t = z}$, respectively: 
\begin{align}
    \tU_{t+1}(\theta)  =  U_{t+1}(\theta) - \mu_t(\oZ_{t-1}, \oX_{t-1}), \quad
    \tind{Z_t = z}  = \indsub{Z_t = z} - e_t^z(\oZ_{t-1}, \oX_{t-1}),
\end{align}
with
\begin{align}
    \mu_t(\oZ_{t-1}, \oX_{t-1}) =\bbE\left\{U_{t+1}(\theta) \mid \oZ_{t-1}, \oX_{t-1}\right\}, \quad
    e_t^z(\oZ_{t-1}, \oX_{t-1}) = \bbP(Z_t = z \mid \oZ_{t-1}, \oX_{t-1}).
\end{align}
\end{theorem}

When the blip functions are restricted to a parametric family, \citet{robins1994correcting} derived the optimal choice of the function $h_t^z(\oZ_{t-1}, \oX_{t-1})$ to achieve the semiparametric efficiency bound. The optimal form is analytically complex. Solving the estimating equations recursively, we can obtain a sequence of blip functions $\gamma_t(\oZ_t, \oX_{t-1})$ for each time $t$. A convenient and widely used choice is to set $h_t^z(\oZ_{t-1}, \oX_{t-1})$ equal to the gradient of the blip function with respect to its parameters:
\begin{align}
    \label{eqn:blip-function-gradient}
    h_t^z(\oZ_{t-1}, \oX_{t-1}) = - \nabla_{\theta} g_t^z(\oZ_{t-1}, \oX_{t-1};\theta).
\end{align}
With this choice, the general estimating equation reduces to the following regression problem:
{\small 
\begin{align}
    \label{eqn:blip-function-regression}
    \min_{\theta} \bbE\Big[\Big\{\tU_{t+1}(\theta) - \sum_{z\in \cZ^{+}} \tind{Z_t = z}  g_t^z(\oZ_{t-1}, \oX_{t-1};\theta)\Big\}^2\Big].\phantom{space} 
\end{align}} 
The equivalence follows directly from the first-order optimality condition of the minimization problem. 

\paragraph{Robustness to outcome misspecification.}
The solution to the estimating equation~\eqref{eqn:blipped-outcome-estimating-equation} is robust to misspecification of outcome models. In randomized experiments with known treatment probabilities, it relaxes any model assumption on the conditional mean of the blipped outcomes. We summarize the robustness property in the following proposition.
\begin{proposition}[Robustness property]
\label{lemma:double-robustness}
    Estimating equations~\eqref{eqn:blipped-outcome-estimating-equation} hold as long as the propensity score $e_t^z(\oZ_{t-1}, \oX_{t-1})$ is correctly specified, allowing for arbitrarily misspecified outcome model $\mu_t(\oZ_{t-1}, \oX_{t-1})$. 
\end{proposition}

In practice, the parametric model assumption on the blip functions may be too restrictive. A natural alternative, motivated by~\eqref{eqn:blip-function-regression}, is to estimate the blip functions using nonparametric or machine learning methods using recursive regression with an $\ell_2$ loss. See \citet{lewis2020double} for discussion.

\begin{remark}
At the end, we add several remarks. First, in the special case of a single time point, our recursive residual learning framework reduces to the residual learner (R-learner) by \citet{nie2021quasi} for HTE estimation. Thus, our approach can be viewed as a longitudinal generalization of the R-learner. Second, while we discuss categorical treatments, the residual learning framework also extends naturally to the continuous treatment setting \citep{zhang2022towards}. 
% Third, our method does not require every unit to have the same horizon. 
\end{remark}

\section{Architecture and training}\label{sec:arch-training}

While theoretically well-grounded, the recursive regression approach faces several practical challenges. First, although being nonparametric, it still requires the specification of a suitable function class for estimating the blip functions. Second, estimation involves many nuisance components, including the propensity score $e_t^z(\oZ_{t-1}, \oX_{t-1})$ and the conditional mean function $\mu_t(\oZ_{t-1}, \oX_{t-1})$. Third, it remains unclear how to efficiently share parameters across different time points in the longitudinal setting. 

To address these issues, we build on the time sequence~\eqref{eqn:data-structure} and propose using a transformer architecture to estimate the blip functions $\gamma_t(\oZ_t, \oX_{t-1})$ for each timepoint $t$. 
Transformers are neural network architectures originally designed for sequential data, and they are particularly well-suited for capturing long-range temporal dependencies, sharing information across time, and flexibly representing high-dimensional function classes.

\subsection{Model architecture}\label{sec:arch}

\paragraph{Sequence encoder.}
The starting point of the architecture is a pair of sequence encoders, which are linear layers used to encode the history of treatment assignments and the feature vector into a latent representation. The fully connected layer weights are shared across time points for both the treatment and feature sequences:
\begin{gather}
    \label{eqn:sequence-encoder}
    \Phi_{\text{Initial}}(Z_t) = \textup{Linear}_Z(Z_t), \quad 
    \Phi_{\text{Initial}}(X_t) = \textup{Linear}_X(X_t).
\end{gather}
These encoders embed treatments and feature vectors into a latent representation but do not capture the temporal order of the inputs. To achieve this, we add positional encodings for the embeddings:
\begin{align}
    \label{eqn:positional-encoding}
    \Phi_{\text{PE}}(Z_t) = \Phi_{\text{Initial}}(Z_t) + \textup{PosEncode}(Z_t), \quad 
    \Phi_{\text{PE}}(X_t) = \Phi_{\text{Initial}}(X_t) + \textup{PosEncode}(X_t),
\end{align}
where $\textup{PosEncode}(t)$ is defined componentwise as follows: for $t = 1, \dots, T$ and $k = 1, \dots, d_{\text{model}}$,
\begin{align}
    \label{eqn:positional-encoding-definition}
    & \textup{PosEncode}(t, k) 
    = 
    \begin{cases}
        \sin(t \cdot 10000^{-k/d_{\text{model}}}), & \text{if } k \text{ is even, } \\
        \cos(t \cdot 10000^{-k/d_{\text{model}}}), & \text{if } k \text{ is odd. }
    \end{cases}
\end{align}
Here $d_{\text{model}}$ is the dimension of the model. 

\begin{figure*}[t]
    \centering
    \includegraphics[width=\textwidth]{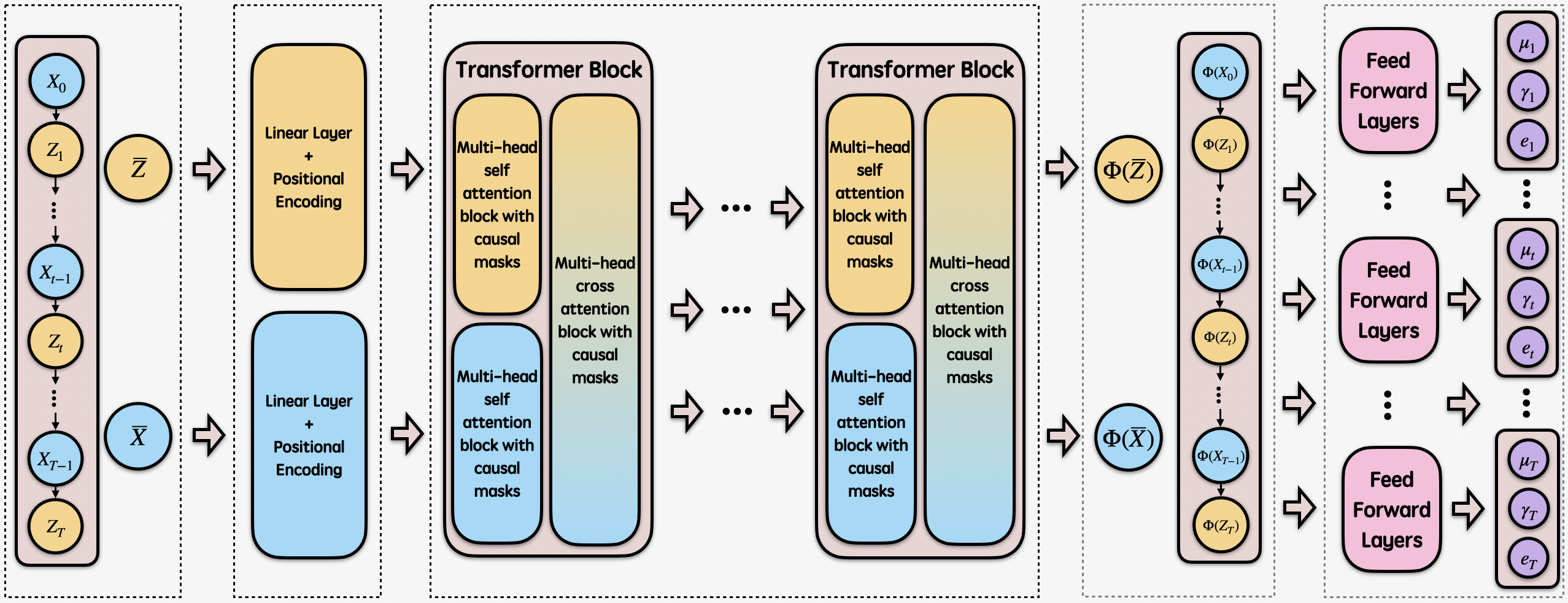}
    \caption{Transformer architecture}
    \label{fig:transformer-architecture}
\end{figure*}

\paragraph{Transformer blocks.}
The encoded sequences are passed through a stack of transformer blocks, as illustrated in Figure~\ref{fig:transformer-architecture}, each designed to capture temporal dependencies and treatment-feature interactions. Each block contains
\begin{itemize}
    \item \textit{Two masked self-attention layers:} one applied to treatment embeddings, another to feature embeddings. The causal mask ensures that earlier time points do not attend to later time points and foresee the future, preserving the longitudinal structure.
    \item \textit{A masked cross-attention layer:} links treatment and feature embeddings by allowing each sequence to attend to the other, again using the causal mask to respect the longitudinal structure.
    \item \textit{Residual connections and normalization:} stabilize learning and prevent vanishing gradients.
\end{itemize}

% Each transformer block is composed of two self-attention layers and a cross-attention layer. One self-attention layer is used to attend to the treatment embeddings, and the other self-attention layer is used to attend to the feature embeddings. The cross-attention layer is used to attend to interleaved information between the treatment and feature embeddings. A causal mask is applied to each attention layer to ensure that earlier time points do not attend to later time points and foresee the future. 

For the transformer block, we use the scaled dot-product attention mechanism with multiple parallel attention heads. More concretely, for a tuple of input $(Q, K, V)$ which represents the query, key, and value matrices, respectively, the dot-product attention is given by:
\begin{align}
    \label{eqn:attention-layer-definition}
    \text{ATTN}(Q, K, V) &= \text{softmax}\left(\frac{QK^{\T}}{\sqrt{d_k}}\right) V.
\end{align}
Here $\text{softmax}$ is the softmax function applied to each row of $QK^{\T}$. The multi-head attention is given by:
\begin{align}
    \label{eqn:multi-head-attention-definition}
    \text{MH}(Q, K, V) & = \text{Concat}(\text{head}_1, \text{head}_2, \dots, \text{head}_h) W^O,
\end{align}
where $\text{head}_i =\text{ATTN}(QW_i^Q, KW_i^K, VW_i^V)$ and $W^O$ is the learned output projection matrix. 

Masking modifies~\eqref{eqn:attention-layer-definition} so that future positions are excluded. Formally, the causal mask is a matrix $M$ with ones in the position where the key is earlier the query and negative infinity vice versa. Let $\odot$ denote the element-wise product. The masked attention is given by: 
\begin{align}
    \label{eqn:attention-layer}
    \text{MaskedATTN}(Q, K, V) &= \text{softmax}\left(\frac{QK^{\T}}{\sqrt{d_k}} \odot M \right) V,
\end{align}
and its multi-head version is
\begin{align}    
    \text{MaskedMH}(Q, K, V) = \text{Concat}(\text{head}_1, \text{head}_2, \dots, \text{head}_h) W^O,
\end{align}
where $\text{head}_i=\text{MaskedATTN}(QW_i^Q,KW_i^K,VW_i^V)$.
After the masked attention, we also apply residual connection and layer normalization; therefore, the final representation after a self-attention layer is given by:
\begin{align}
    \label{eqn:residual-connection}
    \Phi_{\text{ATN}}(\text{Input}) = \text{LayerNorm}(\text{MaskedMH}(\text{Input}) + \text{Input}).
\end{align}
We now revisit the role of the two self-attention layers and the cross-attention layer within each transformer block.

\textit{Two masked self-attention layers:} applied to treatment embeddings and feature embeddings. Both follow the masked attention mechanism in~\eqref{eqn:attention-layer}, with $(Q,K,V)$ constructed from the same sequence.

\textit{A masked cross-attention layer:} in contrast, masked cross-attention couples the treatment and feature sequences. For the treatment subblock, the query $Q$ comes from treatment embeddings while the key-value pair $(K,V)$ comes from feature embeddings. For the feature subblock, the roles are reversed. As in self-attention, a causal mask is applied so that information flows only forward in time.

\noindent\textbf{Output heads.} In the end, there are three sets of output heads, which are feed-forward layers following the transformer blocks: (i) a blip function head, which is generated with a feed-forward layer and encodes the longitudinal HTE; (ii) a propensity score head, which is used to estimate the probability of treatment assignment; (iii) a conditional outcome head, which is used to estimate the conditional means of the outcome. All three heads are trained jointly under the residual learning framework~\eqref{eqn:blip-function-regression}. 

\subsection{Training pipeline}\label{sec:training}

Based on our discussion in Section \ref{sec:methodology}, we summarize our residual training pipeline in Algorithm \ref{alg:terra}. We also highlight several implementation techniques to guarantee reliable training performance.

To stabilize training and ensure reliable convergence, we combine robust optimization and regularization techniques. First, we adopt Adamax as the optimizer, which is a stable variant of Adam \citep{kingma2015adam} that replaces the $\ell_2$ norm with the $\ell_\infty$ norm of the gradients when computing the second moment. This choice provides a more robust estimate of gradient magnitudes and, together with weight decay, can mitigate the influence of extreme gradients that frequently arise in complex architectures. In addition, we apply gradient clipping as an extra safeguard against gradient explosion. To further enhance convergence, we impose a learning rate decay schedule that pushes the learning rate scheme reduction on the training plateau.  

% Let $e$ index epochs and $\eta_0$ be the initial learning rate. Define $k(e)$ as the number of decay events that have occurred up to (and including) epoch $e$. The learning rate at epoch $e$ is $\eta_e \;=\; \eta_0\,\gamma^{\,k(e)}, \ \gamma\in(0,1).$ In practice, if the validation loss fails to improve for $p$ consecutive epochs, we register a decay event and update
% $k(e{+}1)\leftarrow k(e){+}1 \quad\text{and}\quad
% \eta_{e+1}\leftarrow \gamma\,\eta_e
% \;\;\;(\text{equivalently } \eta_{e+1}=\eta_0\gamma^{\,k(e{+}1)}).$ This training strategy can keep steps relatively large early in training and progressively refines the parameters with smaller steps once validation performance plateaus.

To mitigate overfitting, which is particularly pronounced in longitudinal settings, we incorporate standard regularization techniques. To be specific, we adopt dropout within the transformer layers to randomly deactivate units and improve generalization, and early stopping is enforced to terminate training once validation performance deteriorates. 

Finally, to address error propagation across time points, we implement a weighted joint training scheme. Earlier time points are assigned larger weights because they are inherently more difficult to estimate due to accumulated noise. This weighting balances the effective noise variance across time points, yielding more accurate estimates at earlier stages.

Together, these choices yield a training pipeline that is principled, stable, and well-suited to the complexity of longitudinal causal inference with transformers.

\begin{algorithm}[ht!]
\caption{Joint Training for TERRA}
\label{alg:terra}
\DontPrintSemicolon
\SetKwInOut{Input}{Input}\SetKwInOut{Output}{Output}
\Input{Units histories $(\oX_{i,T}, \oZ_{i,T})$ and final outcome $Y_i$; time horizon $T$; treatment set $\cZ$ with control $z^0$; config $\mathcal{C}$ (architecture, loss weights, scheduler, epochs).}
\Output{Parameters $\theta$ of a transformer producing $(\hat e_t^z, \hat\mu_t, \{\hat g_t^z\}_{z\in\cZ^+})$ for $t=1{:}T$.}
\BlankLine
\textbf{Model (per time $t$).} A transformer with causal masks and treatment embeddings outputs: \\
\Indp (i) propensity $\hat e_t^z=\bbP_\theta(Z_t\!=\!z\mid \oZ_{t-1},\oX_{t-1})$;\;
(ii) conditional mean $\hat\mu_t=\bbE_\theta\!\left\{U_{t+1}(\theta)\mid \oZ_{t-1},\oX_{t-1}\right\}$;\;
(iii) blip components $\hat g_t^z(\oZ_{t-1},\oX_{t-1})$ for $z\in\cZ^+$.\;
\Indm
\textbf{Training loop.} Initialize optimizer (e.g., Adamax) and LR scheduler per $\mathcal{C}$; set time weights $w_t$ and loss weights $(\lambda_{\mathrm{prop}},\lambda_{\mathrm{cmu}},\lambda_{\mathrm{blip}})$.\;
\For{epoch $=1$ \KwTo $\mathcal{C}.\texttt{max\_epochs}$}{
  \ForEach{mini-batch $\cB \subset \{1,\dots,N\}$}{
    \tcp{(1) Forward pass and recursive blipping}
    Set $U_{i,T+1} \gets Y_i$ for all $i\in\cB$.\;
    \For{$t=T$ \KwTo $1$}{
      Compute $\hat e_{i,t}^z,\ \hat\mu_{i,t},\ \{\hat g_{i,t}^z\}_{z\neq z^0}$ for all $i\in\cB$.\;
      Define $\hat g_{i,t}^{z^0}\gets 0$ and the realized indicator $I_{i,t}^z\gets \indsub{Z_{i,t}=z}$.\;
      Update blipped outcome: $U_{i,t} \gets U_{i,t+1} - \sum_{z\in\cZ} I_{i,t}^z \hat g_{i,t}^z$.\;
      Residualize: $\ \tilde U_{i,t+1}\gets U_{i,t+1}-\hat\mu_{i,t}$,\quad $\tilde I_{i,t}^z\gets I_{i,t}^z-\hat e_{i,t}^z$.\;
    }
    \tcp{(2) Losses aggregated over time}
    $\mathcal{L}_{\mathrm{prop}}\gets \sum_{t=1}^T w_t\,\mathrm{CrossEntropy}\!\left(\{\hat e_{i,t}^z\}_{z\in\cZ},\,Z_{i,t}\right)$\;
    $\mathcal{L}_{\mathrm{cmu}}\gets \sum_{t=1}^T w_t\,\mathrm{MSE}\!\left(\hat\mu_{i,t},\,U_{i,t+1}\right)$\;
    $\mathcal{L}_{\mathrm{blip}}\gets \sum_{t=1}^T w_t\,\mathrm{MSE}\!\left(\sum_{z\in\cZ}\tilde I_{i,t}^z\,\hat g_{i,t}^z,\ \tilde U_{i,t+1}\right)$\;
    \tcp{(3) Total objective and update}
    $\mathcal{L}\gets \lambda_{\mathrm{prop}}\mathcal{L}_{\mathrm{prop}}+\lambda_{\mathrm{cmu}}\mathcal{L}_{\mathrm{cmu}}+\lambda_{\mathrm{blip}}\mathcal{L}_{\mathrm{blip}}+\mathcal{R}(\theta)$.\;
    Backpropagate $\nabla_\theta\mathcal{L}$; clip gradients; optimizer step; scheduler step.\;
  }
  Apply validation / early stopping and checkpoint the best $\theta$.\;
}
\textbf{Return} best $\theta$ and the learned blip functions $\hat\gamma_t(\oz_t,\oX_{t-1})=\sum_{z\in\cZ}\indsub{z_t=z}\hat g_t^z(\oZ_{t-1},\oX_{t-1})$.\;
\end{algorithm}

\section{Numerical experiments}\label{sec:simulation}
We illustrate the performance of \textsc{TERRA} through two sets of experiments: 
\begin{itemize}
    \item \textit{Simulation studies:} three scenarios capturing both nonlinear and linear blip functions, benchmarked against a set of ML learners based on the residual learning framework.
    \item \textit{Semi-synthetic studies:} comparisons with alternative deep sequence models in a realistic semisythetic baseline to demonstrate the accuracy advantage of TERRA and the potential use case of TERRA in business settings such as marketing attribution. 
%    \item \textbf{[Placeholder for third study]}.
\end{itemize}

Across all experiments, we demonstrate that \textsc{TERRA} always outperforms competing methods and architectures in terms of average MSE, Spearman correlation, and pointwise MSE across time. We trained and evaluated all models on one NVIDIA A100 (80\,GB) GPU. No multi-GPU or distributed training was used.

\subsection{Comparison with ML-based residual learners}
In this subsection, we compare TERRA with other classical CausalML learners \citep{chernozhukov2024applied} in terms of their ability to reconstruct the longitudinal HTE function. In particular, these CausalML learners include LASSO, Bayesian Ridge, Simple linear regression, Multi-layer perceptron (MLP), Random Forest, XGBoost, and Support Vector Regression (SVR). We consider three simulation scenarios with five time-dependent covariates across $T=5$ time points: (i) time-invariant Linear HTE, where the data are generated such that HTE functions across all time points share the same linear form; (ii) time-varying Linear HTE, where the data are generated such that HTE functions across all time points have different linear forms;  and (iii) time-varying nonlinear HTE, where HTE functions across all time points lie in different non-linear forms.

\begin{figure*}[t]
    \centering
    \includegraphics[width=\textwidth]{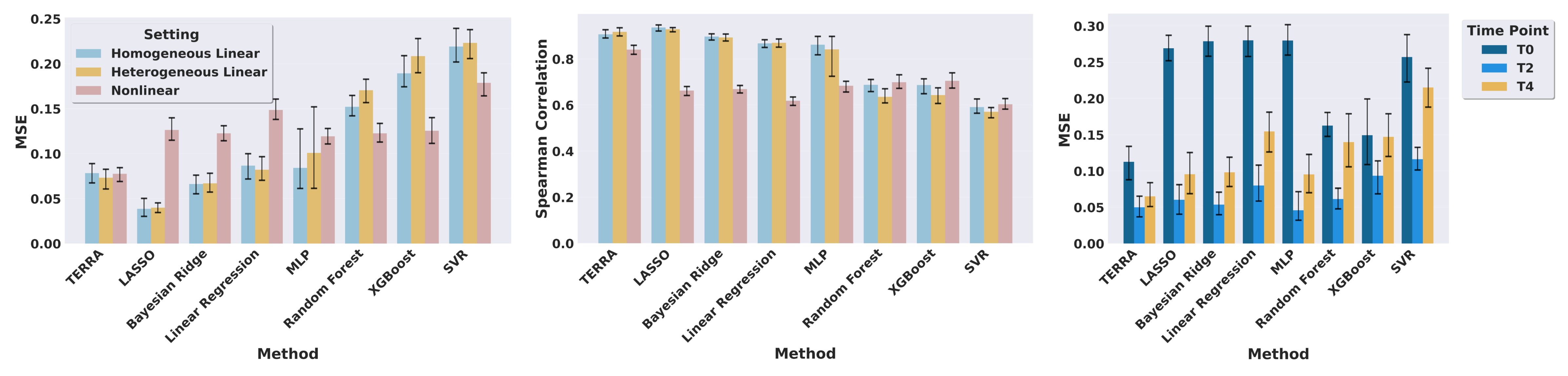}
    \caption{Comparison of different learners under Scenarios 1--3. Results are shown for (a) average MSE, (b) Spearman correlation with the true effects, and (c) MSE at each timepoint.}
    \label{fig:mse}
\end{figure*}

Our results are presented in Figure~\ref{fig:mse}. We report three metrics: (a) average MSE across all timepoints, (b) Spearman Correlation, and (c) MSE at each time point. Across all three scenarios, TERRA consistently outperforms baseline learners. In Figure~\ref{fig:mse}(a), TERRA achieves lower average MSE; in Figure~\ref{fig:mse}(b), it demonstrates stronger rank correlation with the true effects; and in Figure~\ref{fig:mse}(c), it maintains the lowest errors across timepoints, especially at earlier stages where baselines deteriorate. Overall, these results demonstrate that TERRA not only improves average accuracy but also delivers stable, temporally consistent estimates of longitudinal HTE, especially in settings with time-varying heterogeneity.

% \begin{figure}[ht!]
%     \centering
%     \includegraphics[width=1.0\linewidth]{Spearman.png}
%     \caption{Spearman correlation comparison under different settings for different learners. }
%     \label{fig:mse2}
% \end{figure}

% \begin{figure}[ht!]
%     \centering
%     \includegraphics[width=1.0\linewidth]{TimePoints.png}
%     \caption{MSE comparison under different settings for different learners across all time points.}
%     \label{fig:timepoint}
% \end{figure}

\subsection{Semi-synthetic study}\label{sec:iPinYou}

\subsubsection{Setup}
In this section, we present the results of one semi-synthetic study. We create a semi-synthetic data simulator based on a public dataset called iPinYou, which is a popular benchmark dataset for tasks such as bid optimization, click-through rate (CTR) prediction, conversion prediction, and attribution modeling. It was released by the Chinese advertising demand-side platform iPinYou \citep{liao2014ipinyou}.

The iPinYou dataset contained a large volume of user features and advertising logs. We create a synthetic scenario based on the pool of users and advertisements in the Season 3 training data batch of the iPinYou dataset to mimic a setting where a company is randomly displaying a set of Ads to visiting users, and understand the customer journey. This motivation corresponds to the business question of \textit{personalized marketing attribution} \citep{gaur2020attribution}: for a particular user with certain preferences, what is the contribution of each impression along the journey that leads to the final revenue? The concrete data-generating process is described in detail in Section \ref{apdx:iPinYou}, which is a comprehensive and realistic data simulator that allows us to customize the true blip function forms. 

In the following section, we compare TERRA with a set of architectures built for longitudinal causal inference, inspired by \citet{melnychuk2022causal}. The list includes: (i) Causal Transformer (CT) by \citet{melnychuk2022causal}, (ii) Multi-layer perceptron (MLP), which directly uses an MLP to predict the outcome; (iii) Counterfactual Recurrent Network (CRN) by \citet{bica2020estimating}; (iv) G-computation (G-Net) by \citet{li2020g}; (v) Recurrent marginal structural networks (RMSN) by \citet{lim2018forecasting}. 

Besides these results, we also provide additional experiments on the sensitivity of hyperparameters. Due to space limitations, the results are relegated to Section \ref{apdx:hyper}.

\subsubsection{Results}

\begin{figure}[H]
    \centering
    \includegraphics[width=\linewidth]{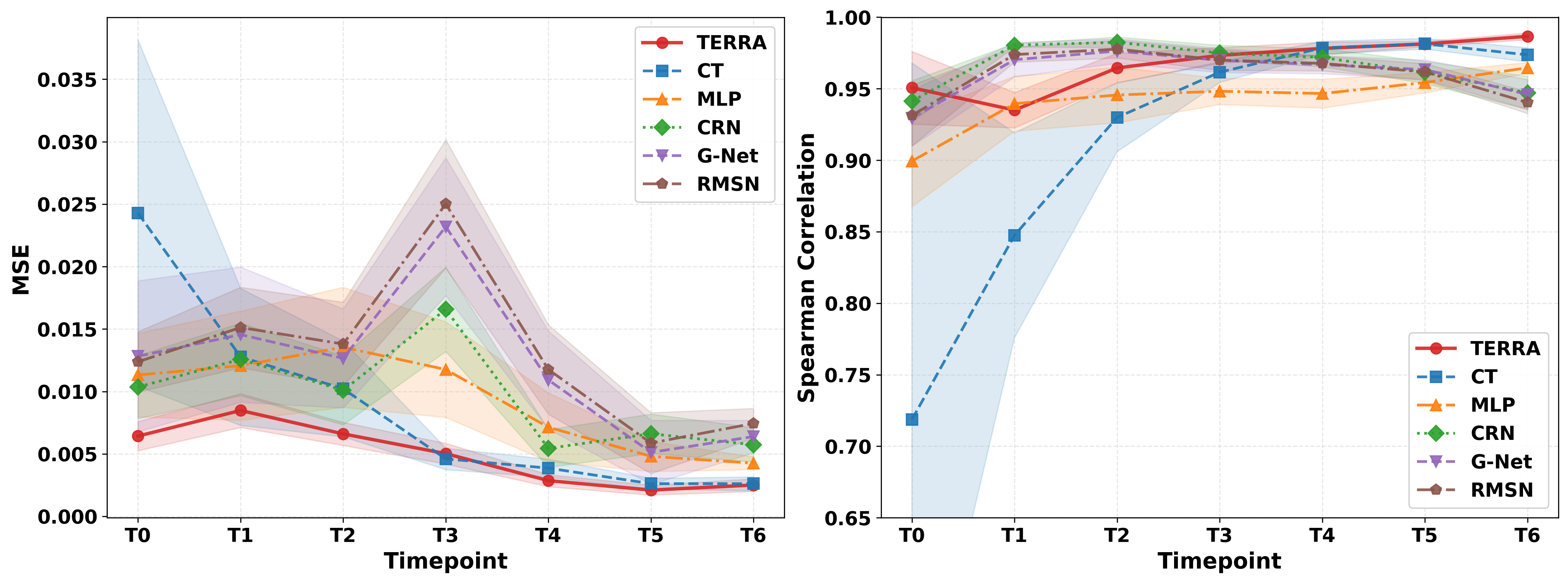}
    \caption{MSE and Spearman correlation comparison across each time point. }
    \label{fig:iPinYou}
\end{figure}

Our results are summarized in Table~\ref{tab:iPinYou} and Figure~\ref{fig:iPinYou}. Across both error and ranking metrics, TERRA delivers the best overall performance, achieving lowest prediction error and near-perfect recovery of the heterogeneous treatment effect ordering. Figure~\ref{fig:iPinYou} further shows that TERRA consistently ranks at top across most timepoints.
\begin{table}[H]
\centering
\caption{Overall Performance Rankings (by MSE and Spearman Correlation)}
\label{tab:iPinYou}
\resizebox{0.47\textwidth}{!}{\begin{tabular}{ccc}
\toprule
\textbf{Method} & \textbf{Overall MSE} & \textbf{Overall Spearman} \\
\midrule
\textbf{TERRA} & \textbf{0.005 $\pm$ 0.000} & \textbf{0.987 $\pm$ 0.001} \\
CT & 0.009 $\pm$ 0.003 & 0.960 $\pm$ 0.015 \\
MLP & 0.009 $\pm$ 0.002 & 0.971 $\pm$ 0.005 \\
CRN & 0.010 $\pm$ 0.002 & 0.978 $\pm$ 0.004 \\
G-Net & 0.012 $\pm$ 0.003 & 0.974 $\pm$ 0.006 \\
RMSN & 0.013 $\pm$ 0.003 & 0.973 $\pm$ 0.004 \\
% MSM & 0.032 $\pm$ 0.002 & 0.756 $\pm$ 0.031 \\
% {R-learner(N)} & 0.159 $\pm$ 0.006 & 0.541 $\pm$ 0.033 \\
% { R-learner(R)} & 0.268 $\pm$ 0.032 & 0.351 $\pm$ 0.061 \\
% Ridge & 0.451 $\pm$ 0.053 & 0.296 $\pm$ 0.052 \\
\bottomrule
\end{tabular}
}
\end{table}
The next tier of architectures (CT, MLP, CRN, G-Net, and RMSN) shows relatively high overall rank correlations but noticeably larger MSEs, suggesting these models capture the relative structure of effects reasonably well yet suffer from calibration error in magnitude. 

Overall, the results indicate that TERRA’s temporal representation and residual/blip modeling better target treatment effect estimation and deliver both accurate magnitudes and reliable rankings. This is a combination especially valuable for policy targeting where ordering and calibration are jointly important.

\section{Discussion}

In this paper, we study longitudinal HTE estimation by extending SNMMs with a recursive residual-learning scheme and integrating it with a Transformer. 

There are many relevant settings not covered in the current paper. First, longitudinal outcomes often face delays or censoring, common in both healthcare and digital experiments \citep{shi2023statistical,shi2024using,bojinov2023design}. Second, beyond estimation, linking temporal HTE to policy learning and decision-making is a natural extension \citep{athey2021policy}. Finally, many applications, such as digital marketing, usually involve network effects, which go beyond the standard no-interference assumption \citep{aronow2017estimating,lu2025design}.

% \subsubsection*{Acknowledgements}
% All acknowledgments go at the end of the paper, including thanks to reviewers who gave useful comments, to colleagues who contributed to the ideas, and to funding agencies and corporate sponsors that provided financial support. 
% To preserve the anonymity, please include acknowledgments \emph{only} in the camera-ready papers. The acknowledgements do not count against the 9-page page limit in the camera-ready.

% \subsubsection*{References}
\newpage
\bibliographystyle{apalike}
\bibliography{ref}

\newpage 
\appendix

\setcounter{theorem}{0}
\renewcommand{\thetheorem}{S\arabic{theorem}}
\renewcommand\theHtheorem{S\thetheorem}

\setcounter{lemma}{0}
\renewcommand{\thelemma}{S\arabic{lemma}}
\renewcommand{\theHlemma}{S\thelemma}

\setcounter{proposition}{0}
\renewcommand{\theproposition}{S\arabic{proposition}}
\renewcommand\theHproposition{S\theproposition}

\renewcommand{\thecorollary}{S\arabic{corollary}}
\setcounter{corollary}{0}
\renewcommand\theHcorollary{S\thecorollary}

\renewcommand{\thedefinition}{S\arabic{definition}}
\setcounter{definition}{0}
\renewcommand\theHdefinition{S\thedefinition}

\renewcommand{\thepage}{S\arabic{page}}
\setcounter{page}{1}

\renewcommand{\theequation}{S\arabic{equation}}
\setcounter{equation}{0}
\renewcommand\theHequation{S\theequation}

\renewcommand{\theassumption}{S\arabic{assumption}}
\setcounter{assumption}{0}
\renewcommand\theHassumption{S\theassumption}

\begin{center}
\Huge 
Supplementary material
\end{center}

Section~\ref{apdx:proof} contains proof of theoretical results. 

Section~\ref{apdx:arch-training} contains additional details on the architecture and training of TERRA, particularly for complexity analysis. 

Section~\ref{apdx:simulation} reports additional simulation setup and results on additional experiments.

\section{Proof of the main results}
\label{apdx:proof}

In this section, we present the detailed proof of the theoretical results in the main text.

\subsection{Proof of Proposition 1}
We prove it by induction. First, \eqref{eqn:blipped-outcome-identity} holds for $t=T$ because for any $z_T$,
\begin{eqnarray*}
    && \bbE\left\{U_{T}(\theta)\mid Z_T=z_T,\oZ_{T-1},\oX_{T-1}\right\} \\
    &=& \bbE\left\{Y- \sum_{z\in \cZ^{+}} \indsub{z_T = z} g_T^z(\oZ_{T-1}, \oX_{T-1};\theta)\mid Z_T=z_T,\oZ_{T-1},\oX_{T-1}\right\} \\
    &=& \bbE\left\{Y(\oZ_{T-1},z_T) - \sum_{z\in \cZ^{+}} \indsub{z_T = z} g_T^z(\oZ_{T-1}, \oX_{T-1};\theta)\mid Z_T=z_T,\oZ_{T-1},\oX_{T-1}\right\} \\
    &=& \bbE\left\{Y(\oZ_{T-1},z_T^0) \mid Z_T=z_T,\oZ_{T-1},\oX_{T-1}\right\}, \\
    &=& \bbE\left\{Y(\oZ_{T-1},z_T^0) \mid \oZ_{T-1},\oX_{T-1}\right\},
\end{eqnarray*}
where the second equality follows from consistency, the third follows from the definition of the blip function $\sum_{z\in \cZ^{+}} \indsub{z_T = z} g_T^z(\cdot)$, and the last follows from the sequential ignorability assumption.
Second, suppose that~\eqref{eqn:blipped-outcome-identity} holds for $t+1$, then it also holds for $t$ because
\begin{eqnarray*}
    \bbE\left\{U_t(\theta)\mid \oZ_{t},\oX_{t-1}\right\} &=& \bbE\left\{U_{t+1}(\theta) - \sum_{z\in \cZ^{+}} \indsub{Z_t = z} g_t^z(\oZ_{t-1}, \oX_{t-1};\theta) \mid \oZ_{t},\oX_{t-1}\right\} \\
    &=& \bbE\left[ \bbE\left\{U_{t+1}(\theta) \mid \oZ_{t},\oX_{t}\right\} \mid \oZ_{t},\oX_{t-1}\right] - \sum_{z\in \cZ^{+}} \indsub{Z_t = z} g_t^z(\oZ_{t-1}, \oX_{t-1};\theta) \\
    &=& \bbE\left[ \bbE\left\{Y(\oZ_{t},\uz_{t+1}^0) \mid \oZ_{t},\oX_{t}\right\} \mid \oZ_{t},\oX_{t-1}\right] - \sum_{z\in \cZ^{+}} \indsub{Z_t = z} g_t^z(\oZ_{t-1}, \oX_{t-1};\theta) \\
    &=& \bbE\left\{Y(\oZ_{t},\uz_{t+1}^0)- \sum_{z\in \cZ^{+}} \indsub{Z_t = z} g_t^z(\oZ_{t-1}, \oX_{t-1};\theta) \mid \oZ_{t},\oX_{t-1}\right\} \\
    &=&  \bbE\left\{Y(\oZ_{t-1},\uz_{t}^0) \mid \oZ_{t},\oX_{t-1}\right\} \\
    &=& \bbE\left\{Y(\oZ_{t-1},\uz_{t}^0) \mid \oZ_{t-1},\oX_{t-1}\right\},
\end{eqnarray*}
where the second equality follows from the law of iterated expectations, the third follows from the induction assumption, the fourth follows from the law of iterated expectations, the fifth follows from the definition of the blip function $\sum_{z\in \cZ^{+}} \indsub{Z_t = z} g_t^z(\cdot)$, and the last follows from the sequential ignorability assumption.

Therefore, by induction, we have equation~\eqref{eqn:blipped-outcome-identity} holds for all $t=1,\ldots,T$.

\subsection{Proof of Theorem 2}
We show the validity of the estimation equation~\eqref{eqn:blipped-outcome-estimating-equation}. First, we introduce the following additional notation. Denote
\begin{align}
    \mu_t(\oZ_{t-1}, \oX_{t-1})
    &= \bbE\left\{U_{t+1}(\theta) \mid \oZ_{t-1}, \oX_{t-1}\right\}, \\
    e_t^z(\oZ_{t-1}, \oX_{t-1})
    &= \bbP(Z_t = z \mid \oZ_{t-1}, \oX_{t-1})
\end{align}
as the conditional mean of the blipped outcome $U_{t+1}(\theta)$ given the history of treatment assignments $\oZ_{t-1}$ and the feature vector $\oX_{t-1}$ and the propensity score, respectively. Under Assumptions~\ref{cond:outcome-consistency}~and~\ref{cond:sequential-randomization},
{\small  
\begin{eqnarray*}
    && \bbE\left[\sum_{z\in \cZ^{+}}\tind{Z_t = z}h_t^z(\oZ_{t-1}, \oX_{t-1}) \left\{\tU_{t+1}(\theta) - \sum_{z\in \cZ^{+}} \tind{Z_t = z} g_t^z(\oZ_{t-1}, \oX_{t-1};\theta)\right\}\right] \\
    % &=& \bbE\left(\bbE\left[\sum_{z\in \cZ^{+}}\tind{Z_t = z}h_t^z(\oZ_{t-1}, \oX_{t-1}) \left\{\tU_{t+1}(\theta) - \sum_{z\in \cZ^{+}} \tind{Z_t = z} g_t^z(\oZ_{t-1}, \oX_{t-1};\theta)\right\}\Big| \oZ_{t},\oX_{t-1}\right]\right) \\
    &=& \bbE\left[\sum_{z\in \cZ^{+}}\tind{Z_t = z}h_t^z(\oZ_{t-1}, \oX_{t-1})\bbE\left\{\tU_{t+1}(\theta) - \sum_{z\in \cZ^{+}} \tind{Z_t = z} g_t^z(\oZ_{t-1}, \oX_{t-1};\theta)\Big| \oZ_{t},\oX_{t-1}\right\}\right] \\
    &=& \bbE\left(\sum_{z\in \cZ^{+}}\tind{Z_t = z}h_t^z(\oZ_{t-1}, \oX_{t-1})\left[\bbE\left\{U_{t+1}(\theta)\Big| \oZ_{t},\oX_{t-1}\right\} -\mu_t(\oZ_{t-1}, \oX_{t-1}) - \sum_{z\in \cZ^{+}} \tind{Z_t = z} g_t^z(\oZ_{t-1}, \oX_{t-1};\theta)\right]\right) \\
    &=& \bbE\left(\sum_{z\in \cZ^{+}}\tind{Z_t = z}h_t^z(\oZ_{t-1}, \oX_{t-1})\left[\bbE\left\{U_{t}(\theta)\Big| \oZ_{t},\oX_{t-1}\right\} -\mu_t(\oZ_{t-1}, \oX_{t-1}) + \sum_{z\in \cZ^{+}} e_t^z(\oZ_{t-1}, \oX_{t-1}) g_t^z(\oZ_{t-1}, \oX_{t-1};\theta)\right]\right) \\
    &=& \bbE\left(\sum_{z\in \cZ^{+}}\tind{Z_t = z}h_t^z(\oZ_{t-1}, \oX_{t-1})\left[\bbE\left\{U_{t}(\theta)\Big| \oZ_{t-1},\oX_{t-1}\right\} +\sum_{z\in \cZ^{+}} e_t^z(\oZ_{t-1}, \oX_{t-1}) g_t^z(\oZ_{t-1}, \oX_{t-1};\theta) -\mu_t(\oZ_{t-1}, \oX_{t-1})\right]\right) \\
    &=& 0,
\end{eqnarray*}
}
where the first equality is by the law of iterated expectations, the fourth equality is by Proposition~\ref{prop:recursive-blipping}, and the last equality follows from the following: Expanding $U_t(\theta)$ in~\eqref{eqn:blipped-outcome-identity} by plugging in the definition of $U_t(\theta)$ in~\eqref{eqn:blipped-outcome-recursive}, we have
\begin{eqnarray*}
    \bbE\left\{U_{t}(\theta)\mid \oZ_{t-1},\oX_{t-1}\right\} &=& 
    \bbE\Big\{U_{t+1}(\theta) - \sum_{z\in \cZ^{+}} \indsub{Z_t = z} g_t^z(\oZ_{t-1}, \oX_{t-1};\theta) \mid \oZ_{t-1}, \oX_{t-1}\Big\} \\
    &=& \mu_{t}(\oZ_{t-1}, \oX_{t-1}) - \sum_{z\in \cZ^{+}} e_t^z(\oZ_{t-1}, \oX_{t-1}) g_t^z(\oZ_{t-1}, \oX_{t-1};\theta),
\end{eqnarray*}

\subsection{Proof of Proposition 3}
In this part, we prove the estimating equation~\eqref{eqn:blipped-outcome-estimating-equation} is valid as long as $e_t^z(\oZ_{t-1}, \oX_{t-1})$ is correctly specified. In other words, we have the following equation for an arbitrary $\mu_t^{\prime}(\oZ_{t-1},\oX_{t-1})$ not necessarily equal to $\mu_t(\oZ_{t-1},\oX_{t-1})$:
\begin{align}
    \bbE\left[\sum_{z\in \cZ^{+}}\tind{Z_t = z}h_t^z(\oZ_{t-1}, \oX_{t-1}) \left\{U_{t+1}(\theta) - \mu_t^{\prime}(\oZ_{t-1},\oX_{t-1}) - \sum_{z\in \cZ^{+}} \tind{Z_t = z} g_t^z(\oZ_{t-1}, \oX_{t-1};\theta)\right\}\right] &= 0. \label{eqn:misspecified_ee}
\end{align}
The left-hand side of~\eqref{eqn:misspecified_ee} is equal to 
{\small 
\begin{eqnarray*}
    && \bbE\left[\sum_{z\in \cZ^{+}}\tind{Z_t = z}h_t^z(\oZ_{t-1}, \oX_{t-1}) \left\{U_{t}(\theta) - \mu_t^{\prime}(\oZ_{t-1},\oX_{t-1}) + \sum_{z\in \cZ^{+}} e_t^z(\oZ_{t-1}, \oX_{t-1}) g_t^z(\oZ_{t-1}, \oX_{t-1};\theta)\right\}\right] \\
    &=& \bbE\left[\sum_{z\in \cZ^{+}}\tind{Z_t = z}h_t^z(\oZ_{t-1}, \oX_{t-1}) \bbE\left\{U_{t}(\theta) - \mu_t^{\prime}(\oZ_{t-1},\oX_{t-1}) + \sum_{z\in \cZ^{+}} e_t^z(\oZ_{t-1}, \oX_{t-1}) g_t^z(\oZ_{t-1}, \oX_{t-1};\theta)\Big| \oZ_t, \oX_{t-1}\right\}\right] \\
    &=& \bbE\left(\sum_{z\in \cZ^{+}}\tind{Z_t = z}h_t^z(\oZ_{t-1}, \oX_{t-1}) \left[\bbE\left\{U_{t}(\theta)\mid \oZ_{t-1}, \oX_{t-1}\right\} - \mu_t^{\prime}(\oZ_{t-1},\oX_{t-1}) + \sum_{z\in \cZ^{+}} e_t^z(\oZ_{t-1}, \oX_{t-1}) g_t^z(\oZ_{t-1}, \oX_{t-1};\theta)\right]\right) \\
    &=& \bbE\left[\sum_{z\in \cZ^{+}}\tind{Z_t = z}h_t^z(\oZ_{t-1}, \oX_{t-1}) \left\{\mu_t(\oZ_{t-1},\oX_{t-1}) - \mu_t^{\prime}(\oZ_{t-1},\oX_{t-1}) \right\}\right] \\
    &=& \bbE\left[\sum_{z\in \cZ^{+}}\left\{\tind{Z_t = z}\mid \oZ_{t-1},\oX_{t-1}\right\}h_t^z(\oZ_{t-1}, \oX_{t-1}) \left\{\mu_t(\oZ_{t-1},\oX_{t-1}) - \mu_t^{\prime}(\oZ_{t-1},\oX_{t-1}) \right\}\right] \\
    &=& 0,
\end{eqnarray*}
}
where the second equality follows from Proposition~\ref{prop:recursive-blipping}, and the last equality follows from the definition of $\tind{Z_t = z}$.

% \begin{align}
%     \label{eqn:blipped-outcome-identity-diagram}
%     \begin{array}{ccccccccccccc}
%         Y & \xrightarrow{\text{blip}} & U_T(\theta) & \xrightarrow{\text{blip}} & \cdots & \xrightarrow{\text{blip}} & U_t(\theta) & \xrightarrow{\text{blip}} & \cdots & \xrightarrow{\text{blip}} & U_2(\theta) & \xrightarrow{\text{blip}} & U_1(\theta) \\
%         \Downarrow & & \Downarrow & & \cdots & & \Downarrow & & \cdots & & \Downarrow & & \Downarrow \\
%         Y(\oZ_T) & \xrightarrow{\text{blip}} & Y(\oZ_{T-1}, \uz_T^0) & \xrightarrow{\text{blip}} & \cdots & \xrightarrow{\text{blip}} & Y(\oZ_{t-1}, \uz_t^0) & \xrightarrow{\text{blip}} & \cdots & \xrightarrow{\text{blip}} & Y(\oZ_1, \uz_2^0) & \xrightarrow{\text{blip}} & Y(\uz_1^0)
%     \end{array}
% \end{align}

\section{Additional details on architecture and training}\label{apdx:arch-training}
\subsection{Complexity analysis}\label{apdx:complexity}

Recall some previous notations. 
Let $N$ be the number of units, $T$ the number of time points, $d$ the hidden width, $L$ the number of transformer layers, $B$ the batch size, $E$ the number of epochs, and $K$ the number of treatment variants.

\paragraph{Per-layer and per-batch costs.}
A standard transformer layer with full (causal) self-attention has time
$O(T^2 d)$ for attention and $O(T d^2)$ for the feed-forward block \citep{vaswani2017attention}.
Thus one layer costs $O(T^2 d + T d^2)$ and $L$ layers cost
\begin{align}
O\!\left(L\,(T^2 d + T d^2)\right)\ \text{per forward/backward pass}.
\end{align}
Task-specific heads (propensity/mean/blip) and the residual/blipping steps add
$O(TK)$--$O(TK d)$ operations per pass, which are lower-order unless $K$ is very large.

\paragraph{Training time.}
With $\lceil N/B\rceil$ batches per epoch and backpropagation (constant factor), the total time over $E$ epochs is
\begin{align}
O\left(E\cdot N\cdot L\cdot (T^2 d + T d^2)\right) .
\end{align}
Inference for a single sequence is $O\!\left(L\,(T^2 d + T d^2)\right)$; with incremental KV
caching, the rollout becomes $O\!\left(L\,(T d + d^2)\right)$ per step (still quadratic overall in $T$ if recomputed) \citep{vaswani2017attention}.

% \paragraph{Memory.}
% Parameter memory is $O(L d^2)$.
% Activation memory during training is dominated by attention maps:
% \[
% \boxed{\ O\!\left(B\,L\,T^2\right)\ \text{(plus }O(BLTd)\text{ for activations)}\ }.
% \]
% This $T^2$ term typically governs the GPU footprint.

\section{Additional experiments}\label{apdx:simulation}

\subsection{Monte Carlo simulation data generating process}\label{apdx:MC}
In this section, we present the details of the Monte Carlo simulation data-generating process (DGP). For each DGP, we take $3,000$ simulated sample based on the following blip function generation pipelines and evaluate TERRA across multiple random seeds. 

\paragraph{Scenario 1: Homogeneous linear blip functions.} In this scenario, all the blip functions take the same linear form: $\gamma_t(\oZ_t, \oX_{t-1}) = Z_t(X_{t-1}^\top \beta + \alpha)$ with $\beta=(0.5,0.3,-0.2,0,0)$ and $\alpha=0.1(t+1)$. The outcomes are generated through:
$$
Y=\sum\limits_{t=1}^T\left\{\gamma_t(\oZ_t, \oX_{t-1})+0.4 \cdot X_{t-1,0}+0.3\cdot X_{t-1,1}\right\}+\varepsilon_1,
$$
where $X_{t-1,0}$ and $X_{t-1,1}$ are the first two covariates at time point $t-1$ and $\varepsilon_1$ is the random noise.

At each time point $t>0$, the treatments are assigned through:
\begin{align*}
    z_1 =0.2X_{t-1,0}-0.1X_{t-1,3} + 0.15 X_{t-1,4} +0.3Z_{t-1},\quad 
    Z_t \sim \text{Bernoulli}(\sigma(z_1)),
\end{align*}
where $\sigma(z)={1}/{(1+e^{-z})}$ is the sigmoid function.

\paragraph{Scenario 2: Heterogeneous linear blip functions.} In this scenario, blip functions still have the linear form but the slopes and the coefficients vary across timepoints: $\gamma_t(\oZ_t, \oX_{t-1}) = Z_t(X_{t-1}^\top \beta(t) + \alpha(t))$. To be specific, 
% \begin{gather*}
%     \beta_0(t)=0.3+0.1t, \quad \beta_1(t)=0.4-0.1t, \quad
%     \beta_2(t)=-0.2 + 0.1 \sin(2 \pi t / T), \\ \beta_3(t)=  0.15 * ((t - (n_timepoints-1)/2) / ((n_timepoints-1)/2) ** 2),\quad \beta_4(t)=0.1 if t < n_timepoints // 2 else -0.1.
% \end{gather*}

\begin{gather}
\beta_0(t) = 0.3 + 0.1\,t, \quad 
\beta_1(t) = 0.4 - 0.1\,t, \quad 
\beta_2(t) = -0.2 + 0.1 \sin\!\left(\frac{2\pi t}{T}\right), \\
\beta_3(t) = 0.15 \left(\frac{t - \frac{T-1}{2}}{\frac{T-1}{2}}\right)^{2}, \quad
\beta_4(t) \quad =
\begin{cases}
0.1, & t < \dfrac{T}{2},\\[4pt]
-0.1, & t \ge \dfrac{T}{2}.
\end{cases}
\end{gather}

The outcomes are generated with the above blip functions and time-varying baseline functions:
{ \begin{align*}
Y=\sum\limits_{t=0}^T\gamma_t(\oZ_t, \oX_{t-1})
+\sum\limits_{t=1}^T\left[\frac{(t+1)}{T}\right](0.3X_{t-1,0}+0.1X_{t-1,1})+\varepsilon_2.
\end{align*}}

The treatment assignment mechanism in scenario 2 uses the same base covariate logits with additional sinusoidal time effect and decaying persistence, i.e.
\begin{align*}
    z_2=0.2X_{t-1,0}-0.1X_{t-1,3} +0.15X_{t-1,4} + z_{2,\mathtt{t}}+ z_{2,\mathtt{p}}Z_{t-1},\quad 
    Z_t\sim \text{Bernoulli}(\sigma(z_2)), 
\end{align*}
where $z_{2,\mathtt{t}}	=0.1\sin(\pi t/(T-1))$ is the time effect and $z_{2,\mathtt{p}}= 0.4-0.05t$ is the persistence.

\paragraph{Scenario 3: Heterogeneous nonlinear blip functions.} In this scenario, blip functions are no longer linear; they have very different nonlinear forms:
\begin{align*}
    &\gamma_0(X_0)=Z_0(0.3 X_{0,0}^2 + 0.2 X_{0,1}^2 + 0.1 X_{0,0} X_{0,1}
    + 0.15 |X_{0,2}| + 0.1),\\
    & \gamma_1(\oZ_1, \oX_0)=Z_1(0.4 \sin(X_{0,0}) + 0.3 \cos(X_{0,1}) + 0.2 X_{0,2} 
    + 0.1 \tanh(X_{0,3}) + 0.2),\\
    & \gamma_2(\oZ_2, \oX_1)=Z_2(0.3 \mathtt{ReLU}(X_{1,0}) + 0.2 \mathtt{ReLU}(X_{1,1}) 
    + 0.1 \exp(\mathtt{clip}(X_{1,2},-2,2)) + 0.15 \log(1+|X_{1,3}|) + 0.3),\\
    &\gamma_3(\oZ_3, \oX_2)=Z_3(0.2 X_{2,0}^3 + 0.15 X_{2,1}^2 X_{2,2} + 0.1 X_{2,0} X_{2,1} X_{2,2} 
    + 0.2 \mathtt{sign}(X_{2,3}) X{2,3}^2 + 0.1 X_{2,4}^2 + 0.4),\\
    & \gamma_4(\oZ_4, \oX_3)= Z_4(0.25 \sin(X_{3,0}^2) + 0.2 \cos(X_{3,1}) X_{3,2} 
    + 0.15 \max(X_{3,3}, X_{3,4}) + 0.1 \min(X_{3,0}^2, 1) + 0.5),
\end{align*}
where $\mathtt{ReLU}(z) = \max(0,z), \ 
\mathtt{sign}(z) = \indsub{z>0} - \indsub{z<0},$ and $
\mathtt{clip}(z,a,b) = \min(\max(z,a),b)$. Here $\indsub{\cdot}$ is the indicator function.

The outcomes are generated through
\begin{align*}
    Y&=\sum\limits_{t=0}^T \gamma_t(\oZ_t,\oX_{t-1})
    + \sum\limits_{t=1}^T
\left[\frac{(t+1)}{T}\right][0.2 \sin(X_{t-1,0}) + 0.1-X_{t-1, 1}^2] + \varepsilon_3.
\end{align*}

The treatment assignment mechanism involves nonlinear covariate transforms, time trend, and nonlinear persistence. In particular, $Z_t\sim\text{Bernoulli}(z_3)$, where 
\begin{align*}
    z_3&=0.3\tanh(X_{t-1,0})+0.2X_{t-1,1}^2+0.15\sin(X_{t-1,2}^2)
    +0.1\mathtt{ReLu}(X_{t-1,3})+0.1\frac{t+1}{T}+0.4 \tanh(2Z_{t-1}-1).
\end{align*}

\subsection{Semi-synthetic data generating process based on the iPinYou dataset}\label{apdx:iPinYou}

In this section, we describe the data-generating process of the semi-synthetic data in Section \ref{sec:iPinYou} in detail. The iPinYou dataset has a wide range of records: (i) user demographic information, such as anonymized age, gender, location, browsing preference, etc., (ii) Ad information, such as Ad slot information (position, size, floor price), creative IDs, etc., (iii) user-ad interactions: impression logs, click logs, and conversion logs. We generate semi-synthetic data with $2.000$ data trajectories from the following pipeline.

\textbf{Feature generation.} For each time step $t$, we construct a $9$-dimensional feature vector $X_t$:
\begin{table}[]
    \centering
    \begin{tabular}{cc}
    \toprule
    Feature & Definition\\
    \midrule 
    $X_1$ & User interest breadth \\
    $X_2$ & Regional demand score \\
    $X_3$ & City socioeconomic score \\
    $X_4$ & Ad slot width normalized\\
    $X_5$ & Ad slot height normalized\\
    $X_6$ & Ad format type\\
    $X_7$ & Ad visibility position\\
    $X_8$ & Log Cumulative Clicks\\
    $X_9$ & Log Cumulative Conversions\\
    \bottomrule
    \end{tabular}
    \caption{Feature dictionary for the semi-synthetic iPinYou dataset}
    \label{tab:placeholder}
\end{table}
At time $0$, the baseline level values are generated from calibrated distributions $\cF_{\text{Emp}, p}~ (p \in [9])$ based on the dataset.; For example, the user interest breadth is sampled from the empirical distribution of the number of user interest tags. Moreover, regional demand and city socioeconomic scores are based on the empirical frequency of user visiting requests from different regions and cities. Temporally, for user attribute features $X_1$ to $X_3$, we keep them fixed. For Ad-relevant features, we apply an AR(1) model with moderate temporal correlation to account for the realistic factor that user behavior might be subject to certain browsing patterns: for $p = 4,5,6,7$,
\begin{align}
    X_{t,p} = 0.6 * X_{t-1,p} + 0.4 * \cF_\text{Emp, p} + \varepsilon.
\end{align}
For cumulative clicks and conversions $X_8$ and $X_9$, we simulate them following a realistic pipeline. Let $p^{\text{click}}_0$ and $p^{\text{conv}}_0$ be the historical click rate and conversion rate simulated from real data, respectively. At each time $t$, the click probability is defined as
\begin{align*}
    p^{\text{click}}_t 
    &= \mathtt{clip}(p^{\text{click}}_0
       + 0.1 \cdot \max\{0, \text{score}_t\}, 0.001, 0.5),
\end{align*}
where $\text{score}_t$ is a function of features and treatment to be defined later in \eqref{eqn:score-t}, showcasing the temporal contribution of the user action, and the click indicator is drawn as
\begin{align}
     \text{clicked}_t \sim \text{Bernoulli}\!\left(p^{\text{click}}_t\right).
\end{align}
Conditional on a click, the conversion probability is defined as follows: if $\text{clicked}_t  = 1$, 
\begin{align*}
    p^{\text{conv}}_t 
    &= \mathtt{clip}\lt(\frac{p^{\text{conv}}_0}
            {p^{\text{click}}_0}
       + 0.05 \cdot \max\{0, \text{score}_t\}, 0.001, 0.3\rt), 
\end{align*}
where $\text{score}_t$ is defined identically as before and 
\begin{align}
    \text{converted}_t &\sim \text{Bernoulli}\!\left(p^{\text{conv}}_t\right).
\end{align}
If $\text{clicked}_t = 0$, then $\text{converted}_t = 0$ deterministically. Finally, the cumulative counters are updated as
by adding these new generations. 

\textbf{Treatment assignment.} For treatment, we can choose a randomization of $K$ creative IDs. In our experiment, we take $K = 4$ with three active treatments and one control. We consider an A/B testing scenario where we fully randomize the creative IDs for each impression, which leads to a propensity score of 0.25 for each level and each impression. 

\textbf{Blip functions. } We generate the blip functions as a linear function of some engineered features derived from the original (note we are not taking a linear blip function from original features to avoid triviality). We construct five engineered features: 
\begin{itemize}
    \item Centered user intensity $\phi(X_1)$: which is a centered version of the number of user tags;
    \item Geo score $\phi(X_2)$: a centered linear combination of region score and city score; 
    \item Format score $\phi(X_3)$: a vector mapping from Ad format to some scores that represent Ad quality;
    \item Visibility scores $\phi(X_4)$: a vector mapping from visibility to some scores that represent Ad visibility quality; 
    \item Past click rates $\phi(X_5)$: portion of clicks within impression logs. 
\end{itemize}

We mimic our simulation under the heterogeneous linear model setting and define the time-varying covariate coefficients as functions of $t$: 
\begin{gather*}
\beta_{1,t} = 0.3 + 0.1 t, \quad 
\beta_{1,t} = 0.4 - 0.05 t, \\
\beta_{1,t} = -0.2 + 0.1 \cdot \sin\!\left(\tfrac{2\pi t}{T}\right), \quad 
\beta_{1,t} = 0.15 \left(\frac{t - T/2}{T/2}\right)^2, \\
\beta_{1,t} = 
    \begin{cases}
      -0.1, & t < T/2, \\
      0.1, & t \geq T/2,
    \end{cases} \quad
\alpha_t = 0.1 (t+1).
\end{gather*}
Moreover, for each treatment arm $k$, we define a treatment scaling factor:
\begin{align}
    c_k = \max\!\left(0.4, \, 1.0 - 0.2 (k-1)\right), \quad k\in \{2,\dots,K\}. 
\end{align}
The treatment blip function for treatment $k$ at time $t$ is then given by
\begin{align*}
\gamma_k(\oZ_t, \oX_t) 
&=c_k \cdot \Bigl(\phi(X_t)^\top \beta_t + \alpha_t
    \Bigr)\cdot \indsub{Z_t = k}.
\end{align*}

\textbf{Outcome generation. } The final realized outcome, which, for example, can be thought of as a revenue generated by the journey, is a compound of temporal contributions from each step:
\begin{align}
    Y = \sum_{t=0}^{T-1} \text{score}_t + \epsilon.
\end{align}
Here, $\text{score}_t$ measures the contribution of step $t$ activities. It is defined as 
\begin{align}\label{eqn:score-t}
    \text{score}_t = \gamma_k(\oZ_t, \oX_t) + \text{baseline}_t, 
\end{align}
where 
\begin{align}
    \text{baseline}_t = \frac{t+1}{T}\cdot \left\{ 0.3 \phi(X_{1,t}) + 0.1 \phi(X_{2,t}) \right\}.
\end{align}

\subsection{Sensitivity to hyperparameter tuning}\label{apdx:hyper}
In this section, we include a set of ablation studies for the choice of hyperparameters. Importantly, we want to understand the impact of the following two sets of parameters: (i) choice of time-specific weights; (ii) choice of the weights for HTE/mean/propensity loss.  

\paragraph{Choice of time-specific weights.} In this study, we compare four weighting strategies, listed in Table \ref{tab:weighting}. From this table, we can see that, they Hyperbolic strategy delivers the best overall trade-off (lowest MSE, highest Spearman). It strongly weights early timepoints but decays smoothly, so the model learns sharp early effects while balancing later ones. Linear decay performs well because it also mildly favors early periods while still preserving substantial weight later, landing close to Hyperbolic. Exponential regime decays too fast, often underweighting mid/late time points and increasing MSE. Uniform is a solid baseline but spreads signal thinly, giving up some performance when early effects carry more information. In practice, we recommend using soft decay regimes such as Hyperbolic or Linear.

\begin{table}[htbp]
\centering
\caption{Performance by weighting strategy with formula $w_t$ (time index $t=0,1,\ldots$).}
\label{tab:weighting}
\setlength{\tabcolsep}{5pt}
\renewcommand{\arraystretch}{1.15}
\begin{tabular}{cccc}
\toprule
\textbf{Strategy} & \textbf{Formula $w(t)$} & \textbf{Overall MSE} & \textbf{Overall Spearman} \\
\midrule
Uniform         & $w_t=1$                                   & 0.0044 & 0.9812 \\
Hyperbolic  & $w_t=\dfrac{10}{t+1}$                     & \textbf{0.0038} & \textbf{0.9883} \\
% hyperbolic\_20  & $w(t)=\dfrac{20}{t+1}$                     & 0.0058 & 0.9845 \\
% hyperbolic\_30  & $w(t)=\dfrac{30}{t+1}$                     & 0.0051 & 0.9854 \\
% exponential\_06 & $w(t)=10\cdot 0.6^{\,t}$                   & 0.0061 & 0.9829 \\
Exponential & $w_t=10\cdot 0.8^{\,t}$                   & 0.0048 & 0.9837 \\
Linear Decay   & $w_t=10\cdot\big(1-0.1\,t\big)$           & 0.0041 & 0.9853 \\
\bottomrule
\end{tabular}
\end{table}

\paragraph{Weight of HTE loss.} In this part, we compare different weight combinations for propensity/mean/HTE loss. We keep $\lambda_{\texttt{prop}} = \lambda_{\texttt{mean}}$, and vary the HTE-focus ratio of $\lambda_{\texttt{HTE}}/\lambda_{\texttt{prop}} = \lambda_{\texttt{HTE}}/\lambda_{\texttt{mean}}$ for performance evaluation. The specifications and results are reported in Table \ref{tab:loss-weights}. Overall, the pattern supports using a medium HTE-centric loss to achieve better performance. 

\begin{table}[ht!]
\centering
\caption{Sensitivity on loss weights}
\label{tab:loss-weights}
\begin{tabular}{ccccc}
\toprule
\textbf{HTE-focus Ratio} & $\lambda_{\texttt{prop}}$ & $\lambda_{\texttt{mean}}$ & $\lambda_{\texttt{HTE}}$ & \textbf{Overall MSE} \\
\midrule
800  & 0.05 & 0.05 & 40.0 & 0.0058 \\
400  & 0.05 & 0.05 & 20.0 & \textbf{0.0048} \\
300    & 0.10 & 0.10 & 30.0 & 0.0052 \\
100     & 0.10 & 0.10 & 10.0 & 0.0055 \\
\bottomrule
\end{tabular}
\end{table}

\end{document}

%% file: CustomizedCommands.tex
\usepackage{etoolbox}
\usepackage{tabularx, booktabs, graphicx}

\makeatletter
\newcommand{\ostar}{\mathbin{\mathpalette\make@circled\star}}
\newcommand{\make@circled}[2]{%
  \ooalign{$\m@th#1\smallbigcirc{#1}$\cr\hidewidth$\m@th#1#2$\hidewidth\cr}%
}
\newcommand{\smallbigcirc}[1]{%
  \vcenter{\hbox{\scalebox{0.77778}{$\m@th#1\bigcirc$}}}%
}
\makeatother

\newcommand{\lt}{\left}
\newcommand{\rt}{\right}

\newcommand{\commenting}[1]{}

\renewcommand{\hat}{\widehat}
\renewcommand{\tilde}{\widetilde}

\newcommand{\E}[1]{{\mathbb{E}\left\{#1\right\}}}

\newcommand{\indep}{\perp \!\!\! \perp}

\ifdef{\see}{\renewcommand{\see}[1]{\text{ (#1)}}}{\newcommand{\see}[1]{\text{ (#1)}}}

\def\boxit#1{\vbox{\hrule\hbox{\vrule\kern6pt\vbox{\kern6pt#1\kern6pt}\kern6pt\vrule}\hrule}}

\newcolumntype{P}[1]{>{\centering\arraybackslash}p{#1}}
\newcolumntype{M}[1]{>{\centering\arraybackslash}m{#1}}
\newcolumntype{L}[1]{>{\raggedright\arraybackslash}m{#1}}

\newcommand{\cB}{{\mathcal{B}}}

\newcommand{\cZ}{{\mathcal{Z}}}

\newcommand{\cF}{{\mathcal{F}}}

\newcommand{\bbP}{{\mathbb{P}}}

\newcommand{\bbE}{{\mathbb{E}}}

\newcommand{\oz}{{\overline{z}}}

\newcommand{\oZ}{{\overline{Z}}}

\newcommand{\oX}{{\overline{X}}}

\newcommand{\uz}{{\underline{z}}}

\newcommand{\tU}{{\tilde{U}}}

%% file: terra-arxiv.bbl
\begin{thebibliography}{}

\bibitem[Aronow and Samii, 2017]{aronow2017estimating}
Aronow, P.~M. and Samii, C. (2017).
\newblock Estimating average causal effects under general interference, with
  application to a social network experiment.
\newblock {\em Annals of Applied Statistics}, 11(4):1912--1947.

\bibitem[Athey and Imbens, 2016]{athey2016recursive}
Athey, S. and Imbens, G. (2016).
\newblock Recursive partitioning for heterogeneous causal effects.
\newblock {\em Proceedings of the National Academy of Sciences},
  113(27):7353--7360.

\bibitem[Athey and Wager, 2021]{athey2021policy}
Athey, S. and Wager, S. (2021).
\newblock Policy learning with observational data.
\newblock {\em Econometrica}, 89(1):133--161.

\bibitem[Battocchi et~al., 2021]{battocchi2021estimating}
Battocchi, K., Dillon, E., Hei, M., Lewis, G., Oprescu, M., and Syrgkanis, V.
  (2021).
\newblock Estimating the long-term effects of novel treatments.
\newblock {\em Advances in Neural Information Processing Systems},
  34:2925--2935.

\bibitem[Berman, 2018]{berman2018beyond}
Berman, R. (2018).
\newblock Beyond the last touch: Attribution in online advertising.
\newblock {\em Marketing Science}, 37(5):771--792.

\bibitem[Bica et~al., 2020]{bica2020estimating}
Bica, I., Alaa, A.~M., Jordon, J., and Van Der~Schaar, M. (2020).
\newblock Estimating counterfactual treatment outcomes over time through
  adversarially balanced representations.
\newblock {\em arXiv preprint arXiv:2002.04083}.

\bibitem[Bojinov et~al., 2023]{bojinov2023design}
Bojinov, I., Simchi-Levi, D., and Zhao, J. (2023).
\newblock Design and analysis of switchback experiments.
\newblock {\em Management Science}, 69(7):3759--3777.

\bibitem[Buhalis and Volchek, 2021]{buhalis2021bridging}
Buhalis, D. and Volchek, K. (2021).
\newblock Bridging marketing theory and big data analytics: The taxonomy of
  marketing attribution.
\newblock {\em International Journal of Information Management}, 56:102253.

\bibitem[Chernozhukov et~al., 2024]{chernozhukov2024applied}
Chernozhukov, V., Hansen, C., Kallus, N., Spindler, M., and Syrgkanis, V.
  (2024).
\newblock Applied causal inference powered by ml and ai.
\newblock {\em arXiv preprint arXiv:2403.02467}.

\bibitem[Gaur and Bharti, 2020]{gaur2020attribution}
Gaur, J. and Bharti, K. (2020).
\newblock Attribution modelling in marketing: Literature review and research
  agenda.
\newblock {\em Academy of Marketing Studies Journal}, 24(4):1--21.

\bibitem[Hu, 2023]{hu2023heterogeneous}
Hu, A. (2023).
\newblock Heterogeneous treatment effects analysis for social scientists: A
  review.
\newblock {\em Social Science Research}, 109:102810.

\bibitem[Imai and Ratkovic, 2013]{imai2013estimating}
Imai, K. and Ratkovic, M. (2013).
\newblock Estimating treatment effect heterogeneity in randomized program
  evaluation.
\newblock {\em The Annals of Applied Statistics}, pages 443--470.

\bibitem[Kennedy, 2023]{kennedy2023towards}
Kennedy, E.~H. (2023).
\newblock Towards optimal doubly robust estimation of heterogeneous causal
  effects.
\newblock {\em Electronic Journal of Statistics}, 17(2):3008--3049.

\bibitem[Kent et~al., 2018]{kent2018personalized}
Kent, D.~M., Steyerberg, E., and Van~Klaveren, D. (2018).
\newblock Personalized evidence based medicine: predictive approaches to
  heterogeneous treatment effects.
\newblock {\em Bmj}, 363.

\bibitem[Kingma and Ba, 2015]{kingma2015adam}
Kingma, D.~P. and Ba, J. (2015).
\newblock Adam: A method for stochastic optimization.
\newblock In {\em Proceedings of the 3rd International Conference on Learning
  Representations (ICLR)}.

\bibitem[K{\"u}nzel et~al., 2019]{kunzel2019metalearners}
K{\"u}nzel, S.~R., Sekhon, J.~S., Bickel, P.~J., and Yu, B. (2019).
\newblock Metalearners for estimating heterogeneous treatment effects using
  machine learning.
\newblock {\em Proceedings of the national academy of sciences},
  116(10):4156--4165.

\bibitem[Lewis and Syrgkanis, 2020]{lewis2020double}
Lewis, G. and Syrgkanis, V. (2020).
\newblock Double/debiased machine learning for dynamic treatment effects via
  g-estimation.
\newblock {\em arXiv preprint arXiv:2002.07285}.

\bibitem[Li et~al., 2020]{li2020g}
Li, R., Shahn, Z., Li, J., Lu, M., Chakraborty, P., Sow, D., Ghalwash, M., and
  Lehman, L.-w.~H. (2020).
\newblock G-net: a deep learning approach to g-computation for counterfactual
  outcome prediction under dynamic treatment regimes.
\newblock {\em arXiv preprint arXiv:2003.10551}.

\bibitem[Liao et~al., 2014]{liao2014ipinyou}
Liao, H., Peng, L., Liu, Z., and Shen, X. (2014).
\newblock ipinyou global rtb bidding algorithm competition dataset.
\newblock In {\em Proceedings of the Eighth International Workshop on Data
  Mining for Online Advertising}, pages 1--6.

\bibitem[Lim, 2018]{lim2018forecasting}
Lim, B. (2018).
\newblock Forecasting treatment responses over time using recurrent marginal
  structural networks.
\newblock {\em Advances in neural information processing systems}, 31.

\bibitem[Lu et~al., 2025]{lu2025design}
Lu, S., Shi, L., Fang, Y., Zhang, W., and Ding, P. (2025).
\newblock Design-based causal inference in bipartite experiments.

\bibitem[Melnychuk et~al., 2022]{melnychuk2022causal}
Melnychuk, V., Frauen, D., and Feuerriegel, S. (2022).
\newblock Causal transformer for estimating counterfactual outcomes.
\newblock In {\em International conference on machine learning}, pages
  15293--15329. PMLR.

\bibitem[Neyman, 1990]{neyman1923application}
Neyman, J. (1923/1990).
\newblock On the application of probability theory to agricultural experiments.
  essay on principles. section 9.
\newblock {\em Statistical Science}, pages 465--472.

\bibitem[Nie and Wager, 2021]{nie2021quasi}
Nie, X. and Wager, S. (2021).
\newblock Quasi-oracle estimation of heterogeneous treatment effects.
\newblock {\em Biometrika}, 108(2):299--319.

\bibitem[Robins, 1986]{robins1986new}
Robins, J. (1986).
\newblock A new approach to causal inference in mortality studies with a
  sustained exposure period—application to control of the healthy worker
  survivor effect.
\newblock {\em Mathematical Modelling}, 7(9-12):1393--1512.

\bibitem[Robins, 1994]{robins1994correcting}
Robins, J.~M. (1994).
\newblock Correcting for non-compliance in randomized trials using structural
  nested mean models.
\newblock {\em Communications in Statistics-Theory and Methods},
  23(8):2379--2412.

\bibitem[Robins, 2004]{robins2004optimal}
Robins, J.~M. (2004).
\newblock Optimal structural nested models for optimal sequential decisions.
\newblock In {\em Proceedings of the Second Seattle Symposium in Biostatistics:
  analysis of correlated data}, pages 189--326. Springer.

\bibitem[Robins et~al., 2000]{robins2000sensitivity}
Robins, J.~M., Rotnitzky, A., and Scharfstein, D.~O. (2000).
\newblock Sensitivity analysis for selection bias and unmeasured confounding in
  missing data and causal inference models.
\newblock In {\em Statistical models in epidemiology, the environment, and
  clinical trials}, pages 1--94. Springer.

\bibitem[Rubin, 1974]{rubin1974estimating}
Rubin, D.~B. (1974).
\newblock Estimating causal effects of treatments in randomized and
  nonrandomized studies.
\newblock {\em Journal of educational Psychology}, 66(5):688.

\bibitem[Shao and Li, 2011]{shao2011data}
Shao, X. and Li, L. (2011).
\newblock Data-driven multi-touch attribution models.
\newblock In {\em Proceedings of the 17th ACM SIGKDD international conference
  on Knowledge discovery and data mining}, pages 258--264.

\bibitem[Shi et~al., 2023]{shi2023statistical}
Shi, L., Wang, J., and Wu, T. (2023).
\newblock Statistical inference on multi-armed bandits with delayed feedback.
\newblock In {\em International Conference on Machine Learning}, pages
  31328--31352. PMLR.

\bibitem[Shi et~al., 2024]{shi2024using}
Shi, L., Wei, W., and Wang, J. (2024).
\newblock Using surrogates in covariate-adjusted response-adaptive
  randomization experiments with delayed outcomes.
\newblock {\em Advances in Neural Information Processing Systems},
  37:71085--71121.

\bibitem[Sun, 2025]{Statsig2025}
Sun, Y. (2025).
\newblock Digital marketing attribution models: A tech survey.
\newblock Accessed: 2025-09-11.

\bibitem[Tran et~al., 2023]{tran2023inferring}
Tran, A., Bibaut, A., and Kallus, N. (2023).
\newblock Inferring the long-term causal effects of long-term treatments from
  short-term experiments.
\newblock {\em arXiv preprint arXiv:2311.08527}.

\bibitem[Vansteelandt and Joffe, 2014]{vansteelandt2014structural}
Vansteelandt, S. and Joffe, M. (2014).
\newblock Structural nested models and g-estimation: the partially realized
  promise.

\bibitem[Vansteelandt and Joffe, 2016]{vansteelandt2016revisiting}
Vansteelandt, S. and Joffe, M. (2016).
\newblock Revisiting structural nested mean models for causal inference with
  survival data.
\newblock {\em Statistical Science}, 31(2):222--240.

\bibitem[Vaswani et~al., 2017]{vaswani2017attention}
Vaswani, A., Shazeer, N., Parmar, N., Uszkoreit, J., Jones, L., Gomez, A.~N.,
  Kaiser, {\L}., and Polosukhin, I. (2017).
\newblock Attention is all you need.
\newblock {\em Advances in Neural Information Processing Systems}, 30.

\bibitem[Wager, 2024]{wager2024causal}
Wager, S. (2024).
\newblock Causal inference: A statistical learning approach.

\bibitem[Xie et~al., 2018]{xie2018false}
Xie, Y., Chen, N., and Shi, X. (2018).
\newblock False discovery rate controlled heterogeneous treatment effect
  detection for online controlled experiments.
\newblock In {\em Proceedings of the 24th ACM SIGKDD international conference
  on knowledge discovery \& data mining}, pages 876--885.

\bibitem[Zhang et~al., 2022]{zhang2022towards}
Zhang, Y., Kong, D., and Yang, S. (2022).
\newblock Towards r-learner of conditional average treatment effects with a
  continuous treatment: T-identification, estimation, and inference.
\newblock {\em arXiv preprint arXiv:2208.00872}.

\end{thebibliography}
